\documentclass[preprint,12pt]{elsarticle}

\usepackage{amssymb}

\journal{Symmetry}

\begin{document}

\begin{frontmatter}

\title{Particle-Hole Transformation in Strongly-Doped Iron-Based Superconductors}

\author{J.P. Rodriguez}


\address{Department of Physics and Astronomy, 
California State University, Los Angeles, California 90032}

\begin{abstract}
An exact particle--hole transformation is discovered in a local-moment model
for a single layer of heavily electron-doped FeSe. 
The model harbors hidden magnetic order between the iron $d_{xz}$ and $d_{yz}$ orbitals
at the wavenumber $(\pi,\pi)$.
It potentially is tied to the magnetic resonances about the very same N\'eel ordering vector
that have been recently discovered in intercalated FeSe.  
Upon~electron doping, 
the local-moment model successfully accounts for the electron-pocket Fermi surfaces 
observed experimentally
at the corner of the two-iron Brillouin zone in electron-doped FeSe, 
as well as for isotropic Cooper pairs.
Application of the particle--hole transformation predicts a
 surface-layer iron-based superconductor at strong hole doping 
that exhibits high $T_c$, and  that shows hole-type Fermi-surface pockets at the center of the two-iron
Brillouin zone.
\end{abstract}



\end{frontmatter}


\section{Introduction}
The discovery of iron-based superconductors has identified a new route
in the search for high critical temperatures~\cite{new_sc}.
Iron atoms in these materials lie in 
weakly coupled stacks of square lattices~\cite{paglione_greene_10}.
Electronic conduction resides within such layers,
where charge carriers are primarily electrons/holes from iron $3d$ levels.
The optimum critical temperature in iron-pnictide materials, in particular,
coincides with imperfect nesting between hole Fermi-surface pockets
at the center  of the Brillouin zone 
and electron Fermi-surface pockets
at momenta along the principal axes of the square lattice of iron atoms
that coincide with commensurate spin-density wave (cSDW) order.
Strong hole doping can destroy such nesting.
In particular,
angle-resolved photoemission spectroscopy (ARPES)
finds that
the electron bands at cSDW momenta
rise completely above the Fermi level
in the series of compounds
(Ba$_{1-x}$K$_x$)Fe$_2$As$_2$ at $0.5 < x < 0.7$~\cite{malaeb_prb_12}.
ARPES on the end-member of the series KFe$_2$As$_2$,
with superconducting $T_c \cong 4$ K,
reveals only hole Fermi surface pockets~\cite{sato_09}.
Density-functional-theory calculations recover the Lifshitz transtion at which the electron-type
Fermi surface pockets disappear, but at a larger critical concentration
of hole doping~\cite{khan_johnson_14}, $x_c = 0.9$.

Strong electron doping can also destroy nesting in iron-based superconductors.
ARPES on a monolayer of FeSe over a doped SrTiO$_3$ (STO) substrate
and ARPES on intercalated FeSe find only  
electron Fermi surface pockets at cSDW momenta~\cite{peng_14,lee_14,zhao_16}.
Hole bands at the center of the Brillouin zone
lie buried below the Fermi level.
Unlike heavily hole-doped compounds such as KFe$_2$As$_2$, however,
the FeSe surface layer shows high 
critical temperatures, $T_c\sim 100$ K, for superconductivity~\cite{ge_15}.
In addition, ARPES~\cite{peng_14,lee_14,zhao_16}
and scanning tunneling microscopy (STM)~\cite{fan_15,yan_16}
on such surface layers of FeSe
find evidence for an isotropic gap over the electron Fermi surface pockets,
with no nodes.
Finally,    
a Mott insulator phase is reported nearby at low electron doping
in single-layer FeSe/STO and
in voltage-gate tuned thin films of FeSe~\cite{zhou_14,hosono_16}.
In contrast to itinerant magnetism,
which is driven by Fermi-surface nesting,
and which has some success in describing superconductivity in iron-pnictide materials~\cite{HKM_11,Chubukov_15},
the limit of strong on-site electron repulsion~\cite{Si&A,jpr_ehr_09}
may then be a valid starting point to describe superconductivity
in heavily electron-doped FeSe.

Below, we identify a particle--hole transformation 
for a local-moment description of
a single layer in an iron-based superconductor~\cite{Si&A,jpr_ehr_09,jpr_10}
that includes the minimum 
$d_{xz}$ and $d_{yz}$ iron orbitals~\cite{jpr_mana_pds_11,jpr_mana_pds_14,jpr_16}.
At half filling of electrons,
a doped Mott insulator
results  in the limit of strong iron-site Coulomb repulsion~\cite{jpr_10}.
Above half filling (electron doping),
mean field and exact calculations based on
a hidden  half metal state
predict electronic structure that is very similar to that
shown by heavily electron-doped (high-$T_c$)
surface layers of FeSe~\cite{jpr_17,jpr_rm_18}.
The exact calculations
also predict isotropic Cooper pairs at the electron Fermi surface pockets,
in addition to remnant isotropic Cooper pairs of opposite sign on buried hole bands.
Application of the particle--hole transformation to a surface layer of FeSe
predicts a surface-layer iron-based superconductor that is heavily hole-doped,
and that exhibits high $T_c$~\cite{jpr_16}.

\section{Local-Moment Hamiltonian}
Our starting point is a
two-orbital $t$-$J$ model over the square lattice,
where intra-orbital on-site Coulomb repulsion
is  strong~\cite{Si&A,jpr_ehr_09,jpr_mana_pds_11,jpr_mana_pds_14,dagotto_11,dagotto_12}:
%
\begin{equation}
\begin{array}{ll}
H = & \sum\limits_{\langle i,j \rangle}{[-(t_1^{\alpha,\beta} c_{i, \alpha,s}^{\dagger} c_{j,\beta,s} + {\rm h.c.}) + J_1^{\alpha,\beta} {\bf S}_{i, \alpha} \cdot {\bf S}_{j, \beta}]} + \\
 & \sum\limits_{\langle\langle i,j \rangle\rangle}{[-(t_2^{\alpha,\beta} c_{i, \alpha,s}^{\dagger} c_{j,\beta,s} + {\rm h.c.}) + J_2^{\alpha,\beta} {\bf S}_{i, \alpha} \cdot {\bf S}_{j, \beta}]} + \\
 & \sum\limits_{i}{(J_0 {\bf S}_{i, d-}\cdot {\bf S}_{i, d+} +  U_0^{\prime} {\bar n}_{i,d+} {\bar n}_{i,d-} + {\rm lim}_{U_0\rightarrow\infty} U_0 n_{i,\alpha,\uparrow} n_{i,\alpha,\downarrow}\bigr)}.
\end{array}
\label{tJ}
\end{equation}
Above, ${\bf S}_{i,\alpha}$ is the spin operator that acts on spin $s_0 = 1/2$ states
of $d- = d_{(x-iy)z}$ and $d+ = d_{(x+iy)z}$ orbitals $\alpha$
in iron atoms at site $i$.
Repeated orbital and spin indices
in Equation~(\ref{tJ})
are summed over.
Nearest neighbor and next-nearest neighbor Heisenberg exchange
across the respective links $\langle i,j\rangle$ and $\langle\langle i,j\rangle\rangle$
is controlled by
the coupling constants
$J_1^{\alpha,\beta}$ and $J_2^{\alpha,\beta}$.
They are due primarily to super exchange~\cite{Si&A}.
Hopping of an electron in orbital $\alpha$ to an unoccupied
neighboring orbital $\beta$ is controlled by
the matrix elements $t_1^{\alpha,\beta}$ and $t_2^{\alpha,\beta}$.
Finally, $J_0$ is a ferromagnetic exchange coupling constant
that imposes Hund's Rule.
The last term in Equation~(\ref{tJ}) 
suppresses double occupancy at a site-orbital,
where
$n_{i,\alpha,s} = c_{i, \alpha,s}^{\dagger} c_{i, \alpha,s}$
is the occupation operator for
a spin-$s$ electron in orbital $\alpha$ at site $i$.
The next-to-last term in Equation~(\ref{tJ})
measures the energy cost, $U_0^{\prime} > 0$, 
of a pair of holes at an iron site, 
whereas ${\bar n}_{i,\alpha} = 1-\sum_s n_{i,\alpha,s}$
counts holes at site-orbitals below half filling.
Observe that ${\bar n}_{i,\alpha}$ 
can be replaced by
$-{\bar n}_{i,\alpha}$,
which counts singlet pairs at site-orbitals above half filling.
Finally,    
notice that the operation
$d\pm\rightarrow e^{\pm i\theta} d\pm$ is equivalent to
a rotation of the orbitals by an angle $\theta$ about the $z$ axis.
Spin and occupation operators remain invariant under it.
Magnetism described by
the two-orbital $t$-$J$ model in Equation~(\ref{tJ}) is hence isotropic,
which suppresses orbital order 
and nematicity~\cite{xu_muller_sachdev_08,yoshizawa_simayi_12}.

Because the spin-1/2 moments live on isotropic $d\pm$ orbitals,
two isotropic nearest neighbor 
and next-nearest neighbor 
Heisenberg exchange coupling constants exist:
\begin{equation}
J_n^{\parallel} = J_n^{d\pm,d\pm}  
\quad {\rm and} \quad
J_n^{\perp} = J_n^{d\pm,d\mp}
\quad (n=1,2). \nonumber
\end{equation}
The isotropy of the $d\pm$ orbitals also implies intra-orbital hopping matrix
elements that are isotropic and real:
$t_n^{\parallel} = t_n^{d\pm,d\pm}$ for $n=1,2$.
Finally, the reflection properties of the $d_{xz}$ and $d_{yz}$ orbitals
also imply real inter-orbital hopping matrix elements
between nearest neighbors,
with $d$-wave symmetry~\cite{jpr_mana_pds_14}:
$t_1^{\perp} ({\bf{\hat y}}) = -t_1^{\perp} ({\bf{\hat x}})$,
where $t_1^{\perp} = t_1^{d\pm,d\mp}$.
Inter-orbital next-nearest neighbor hopping matrix elements
$t_2^{d\pm,d\mp}$ also show $d$-wave symmetry,
but they are pure imaginary.
They consequently result in hybridization
of the $d_{xz}$ and $d_{yz}$ orbital bands.
Table~\ref{table0} summarizes the expected phase diagram of the
two-orbital $t$-$J$ model in Equation~(\ref{tJ})
near a quantum critical point
into hidden magnetic order~\cite{jpr_10,jpr_17,jpr_16}.

\begin{table}
\centering
\caption{Groundstate of two-orbital $t$-$J$ model in Equation~(\ref{tJ})
in the presence of magnetic frustration:
$J_2^{\parallel}+J_2^{\perp} > {1\over 2}(J_1^{\parallel}+J_1^{\perp})$.
Hund coupling is tuned to the QCP at half filling,
$\Delta_{cSDW}\rightarrow 0$ (Figure~\ref{spin_waves}),
which separates a cSDW when it's strong 
from hidden magnetic order when it is weak (Ref.~\cite{jpr_10}).
Captions to Figures~\ref{spin_waves} and \ref{no_hunds_rule} give example model parameters.}
\label{table0}
\begin{tabular}{ccc}
\hline
Filling, Bands &
$J_1^{\parallel} < J_1^{\perp}$ & $J_1^{\parallel} > J_1^{\perp}$ \\
\hline
half filling, none   &
hidden ferromagnet & hidden N\'eel \\
hole dope, hole bands @ $\Gamma$ &
hidden half metal, FS @ $\Gamma$ & nested cSDW metal? \\
$e^{-}$ dope, $e^{-}$ bands @ M & nested cSDW metal? & hidden half metal, FS @ M \\
\hline
\end{tabular}
\end{table}

\begin{figure}
\centering
\includegraphics{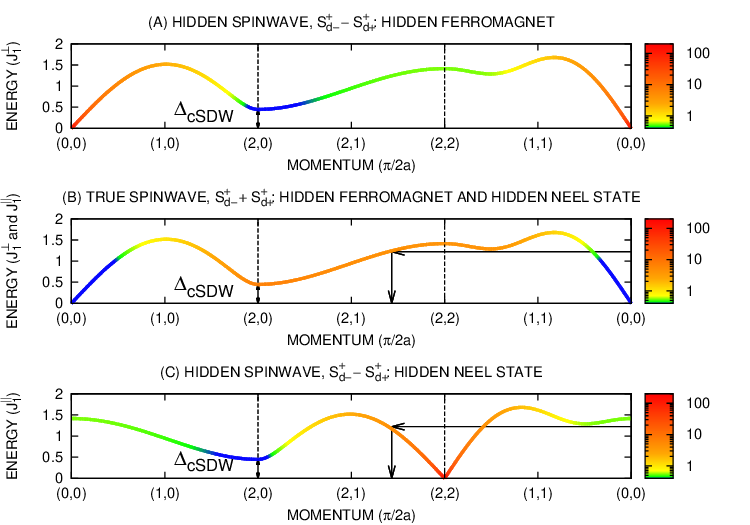}
\caption{Linear spin-wave spectrum (Ref.~\cite{jpr_10})
for (\textbf{a},\textbf{b}) hidden ferromagnet at Heisenberg coupling parameters
$J_1^{\parallel} = 0$, $J_1^{\perp} > 0$, and
$J_2^{\parallel} = 0.3\, J_1^{\perp} = J_2^{\perp}$,
at Hund coupling $-J_0 = -J_{0c} - 0.1 J_1^{\perp}$.
Here, $-J_{0c}$ is the critical Hund coupling
at which $\Delta_{cSDW}\rightarrow 0$.
Model parameters become
(\textbf{b},\textbf{c})
$J_1^{\parallel} > 0$, $J_1^{\perp} = 0$, 
$J_2^{\parallel} = 0.3\, J_1^{\parallel} = J_2^{\perp}$,
and $-J_0 = -J_{0c} - 0.1 J_1^{\parallel}$
in the  hidden N\'eel state after application
of the particle--hole transformation.
Color code represents spectral weight.}
\label{spin_waves}
\end{figure}

\begin{figure}
\centering
\includegraphics{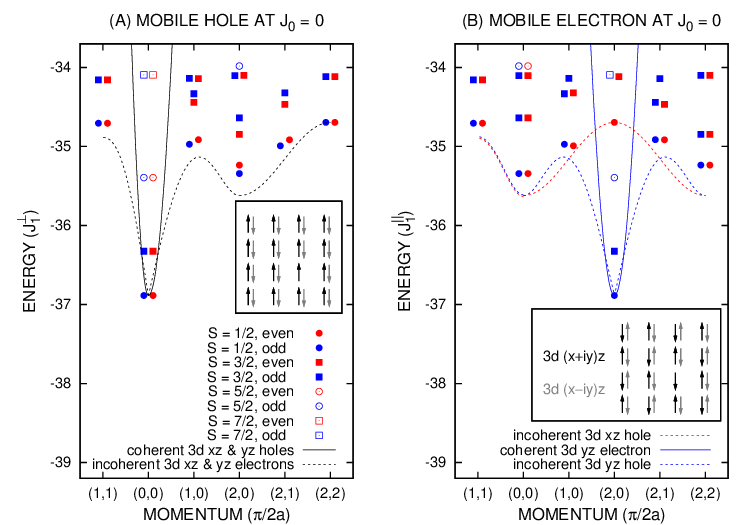}
\caption{Exact spectra for $t$-$J$ model (Equation~(\ref{tJ})), 
over a periodic $4\times 4$ lattice,
with hopping parameters (\textbf{a}) 
$t_1^{\parallel} = -3\, J_1^{\perp}$, 
$t_1^{\perp}({\bf{\hat x}}) = -2\, J_1^{\perp}$,
$t_1^{\perp}({\bf{\hat y}}) = +2\, J_1^{\perp}$,
$t_2^{\parallel} = -\, J_1^{\perp}$, and $t_2^{d\pm,d\mp} = 0$
in the mobile-hole case ($31$~electrons).
Model parameters transform to (\textbf{b})
$t_1^{\parallel} = 2\, J_1^{\parallel}$, 
$t_1^{\perp}({\bf{\hat x}}) = +3\, J_1^{\parallel}$,
$t_1^{\perp}({\bf{\hat y}}) = -3\, J_1^{\parallel}$, 
$t_2^{\parallel} = -\, J_1^{\parallel}$,
and $t_2^{d\pm,d\mp} = 0$
in the mobile-electron case ($33$ electrons).
Heisenberg exchange coupling constants
are given in the caption to Figure~\ref{spin_waves}.
Even/odd quantum number refers to parity under $P_{d,{\bar d}}$.}
\label{no_hunds_rule}
\end{figure}

\section{Particle--Hole Transformation}
\begin{figure}
\centering
\includegraphics{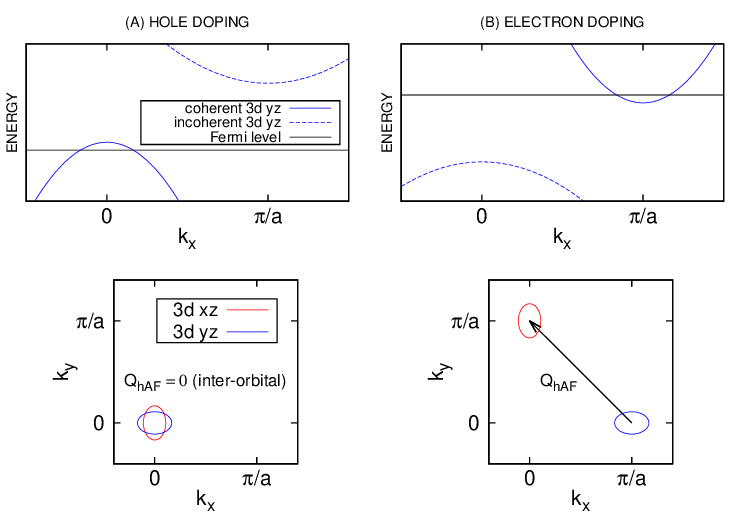}
\caption{Electronic structure of half metal states characterized by 
hidden (inter-orbital) magnetic order 
at wavenumber ${\bf Q}_{\rm hAF}$ (see insets to Figure~\ref{no_hunds_rule}).
Dispersions in energy are fixed at
wavenumber $k_y = 0$.}
\label{e_strctr}
\end{figure}

As   shown below,
the bipartite nature of the square lattice
of iron atoms that stacks up to form 
iron-based superconductors 
allows us to define the following  particle--hole transformation 
in momentum space 
for electrons in either the $d_{xz}$ or  $d_{yz}$ orbitals.
The corresponding electron destruction operator reads
\begin{equation}
c_s(k_0,{\bf k}) = {\cal N}^{-1/2}\sum_{\alpha = 0}^1\sum_i 
e^{-i(k_0 \alpha + {\bf k}\cdot{\bf r}_i)} c_{i,\alpha,s},
\label{mmntm_spc}
\end{equation}
where
${\cal N} = 2 N_{\rm Fe}$ denotes the number of sites-orbitals
on the square lattice of iron atoms,
and where the indices $0$ and $1$ denote
the $d-$ and $d+$ orbitals $\alpha$.
The quantum numbers $k_0 = 0$ and $\pi$ therefore represent
the $d_{xz}$ and the $(-i)d_{yz}$ orbitals.
We then define the particle--hole transformation by the~replacements
\begin{equation}
c_s(k_0,{\bf k})\rightarrow c_s^{\dagger}(k_0,{\bf k}+{\bf Q}_{k_0})
\quad {\rm and} \quad
c_s^{\dagger}(k_0,{\bf k}) \rightarrow c_s(k_0,{\bf k}+{\bf Q}_{k_0}),
\label{p_to_h}
\end{equation}
where ${\bf Q}_{0} = (\pi/a){\bf{\hat y}}$ and ${\bf Q}_{\pi} = (\pi/a){\bf{\hat x}}$.
Figure~\ref{e_strctr} displays the action of 
the above transformation on electronic structure: (A) $\leftrightarrow$ (B).
What then is the form of the 
above particle--hole transformation in real space 
for electrons in $d\pm$ orbitals?
Comparison of Equations~(\ref{mmntm_spc}) and (\ref{p_to_h}) yields the 
equivalent particle--hole transformation in real space:
\begin{equation}
c_{i,\alpha,s}\rightarrow (-1)^{y_i/a} c_{i,p_i(\alpha),s}^{\dagger}
\quad {\rm and} \quad
c_{i,\alpha,s}^{\dagger}\rightarrow (-1)^{y_i/a} c_{i,p_i(\alpha),s}
\label{spc}
\end{equation}
%
where $p_i(d\pm) = d\pm$ for iron sites $i$ on the $A$ sublattice of the checkerboard,
and where $p_i(d\pm) = d\mp$ for iron sites $i$ on the $B$ sublattice of the checkerboard.
(see \ref{app} for details.)

It is useful now to note that application of the particle--hole transformation in Equation~(\ref{spc}) results in a new next-nearest neighbor inter-orbital
hopping matrix element ${\bar t}_2^{d\pm,d\mp}$ that is pure imaginary,
but that alternates in sign between the $A$ versus the $B$ sites of the checkerboard.  
It does not describe mixing of the $d_{xz}$ and $d_{yz}$ orbitals in iron-based superconductors~\cite{jpr_mana_pds_14},
and thus we turn it off
entirely: $t_2^{d\pm,d\mp} = 0$.
The two-orbital $t$-$J$ model Hamiltonian in Equation~(\ref{tJ})
now maintains its form
after making the replacements Equation~(\ref{spc}).
Nearest neighbor model parameters, however, transform to
%
\begin{eqnarray}
{\bar J}_1^{\,\parallel} = J_1^{\perp}
\quad &{\rm and}& \quad
{\bar J}_1^{\,\perp} = J_1^{\parallel}, \nonumber \\
{\bar t}_1^{\,\parallel} = -t_1^{\perp}({\bf{\hat x}})
\quad &{\rm and}& \quad
{\bar t}_1^{\,\perp}({\bf{\hat x}}) = -t_1^{\parallel},
\label{tJbar1}
\end{eqnarray}
%
with ${\bar t}_1^{\,\perp}({\bf{\hat y}}) = -{\bar t}_1^{\,\perp}({\bf{\hat x}})$.
Next-nearest neighbor model parameters
$t_2^{\parallel}$, $J_2^{\parallel}$ and $J_2^{\perp}$
remain unchanged.
Finally,    
on-site parameters $J_0$ for ferromagnetic Hund coupling, 
$U_0$ for intra-orbital Coulomb repulsion,  
and $U_0^{\prime}$ for inter-orbital Coulomb repulsion
also remain unchanged.
Here,
the occupation operators $n_{i,\alpha,s}$ in the divergent
Hubbard term must be replaced by $1-n_{i,\alpha,s}$.

\section{Half Filling}
Consider half filling, with a density of electrons equivalent to one electron per site, per $d\pm$ orbital.
No hopping of electron is then possible in the present limit, $U_0\rightarrow\infty$.
It results in the Heisenberg model associated with 
the coupling constants $J_0$, $J_n^{\parallel}$ and $J_n^{\perp}$ in the model Hamiltonian in Equation~(\ref{tJ}).
The order parameter for hidden magnetic order at wavenumber ${\bf Q}$ is defined by
$O_{hAF} = \sum_i \langle S_{i,d-}^{+} - S_{i,d+}^{+}\rangle\, {\rm exp}(i{\bf Q}\cdot{\bf r}_i)$,
which is equal to
\begin{equation}
O_{hAF} =
 \hbar  \sum_{\bf k} \langle
{c}_{\uparrow}^{\dagger} (0,{\bf k}+{\bf Q}) {c}_{\downarrow}(\pi,{\bf k})
+{c}_{\uparrow}^{\dagger} (\pi,{\bf k}+{\bf Q}) {c}_{\downarrow}(0,{\bf k})\rangle .
\label{hidden_op}
\end{equation}
%
Notice that such hidden magnetic order is equivalent to spin-density-wave order
{\it between} the $d_{xz}$ and $d_{yz}$ orbitals~\cite{jpr_16,jpr_rm_18}.
Application of the particle--hole transformation in Equation~(\ref{p_to_h})
yields minus the complex conjugate of Equation~(\ref{hidden_op})
plus the replacement
${\bf Q}\rightarrow {\bf Q}+(\pi/a)({\bf{\hat x}}+{\bf{\hat y}})$.
In the presence of off-diagonal frustration,
at weak enough Hund coupling,
linear spin--wave theory
applied to the resulting Heisenberg model in Equation~(\ref{tJ})
finds such long-range order
at ${\bf Q} = 0$~\cite{jpr_10}.
It corresponds to hidden ferromagnetic order~\cite{jpr_16}:
$\nwarrow_{d-}\searrow_{d+}$ (see the inset to Figure~\ref{no_hunds_rule}a).
Figure~\ref{spin_waves}a,b
shows the corresponding spin-wave spectra for the hidden ferromagnet
at Heisenberg coupling constants
$J_1^{\parallel} = 0$, $J_1^{\perp} > 0$,
and $J_2^{\parallel} = 0.3\, J_1^{\perp} = J_2^{\perp}$,
at sub-critical Hund coupling characterized by
a spin gap $\Delta_{cSDW} > 0$ at cSDW wavenumbers.
The cSDW spin gap closes at a critical Hund's Rule coupling constant of
$-J_{0c} = 2(J_1^{\perp}-J_1^{\parallel})-4J_2^{\parallel}$
in such case~\cite{jpr_10,jpr_mana_pds_11}.
Following the particle--hole transformation in Equation~(\ref{tJbar1}) of the model parameters,
Figure~\ref{spin_waves}b,c
shows the spin--wave spectra
for  Heisenberg coupling constants
$J_1^{\parallel} > 0$, $J_1^{\perp} = 0$,
and $J_2^{\parallel} = 0.3\, J_1^{\parallel} = J_2^{\perp}$,
but with $J_0$ unchanged.
(The critical Hund's Rule coupling constant is now
$-J_{0c} = 2(J_1^{\parallel}-J_1^{\perp})-4J_2^{\parallel}$.)
Figure~\ref{spin_waves}c displays a Goldstone mode at wavenumber ${\bf Q} = (\pi/a)({\bf{\hat x}}+{\bf{\hat y}})$,
which is evidence for a  hidden N\'eel state.
This hidden antiferromagnet shows opposing N\'eel order per $d\pm$ orbital
(see the inset to Figure~\ref{no_hunds_rule}b and Ref.~\cite{jpr_17}),
which is consistent with the particle--hole transformation in Equation~(\ref{spc})
of the hidden ferromagnet.
Notice that the spectrum of hidden spin--waves (Figure~\ref{spin_waves}c)
is obtained by shifting the spectrum of
its particle--hole conjugate (Figure~\ref{spin_waves}a) 
by the wavenumber $(\pi/a)({\bf{\hat x}}+{\bf{\hat y}})$~\cite{jpr_10}.

True spin--wave and hidden spin--wave excitations,
${\bf S}_{d-}+{\bf S}_{d+}$ and
${\bf S}_{d-}-{\bf S}_{d+}$,
are, respectively, even and odd under orbital swap, $P_{d,{\bar d}}$.
Turning on hopping of electrons $t_2^{d\pm,d\mp}$ in Equation~(\ref{tJ})
that is pure imaginary, with $d$-wave symmetry,
hybridizes the $d_{xz}$ and $d_{yz}$ orbitals~\cite{jpr_mana_pds_14},
which breaks this symmetry away from half filling.
It will mix true and hidden spin--waves,
especially when they are degenerate.  
The~arrows in Figure~\ref{spin_waves}b,c
for spectra in the hidden N\'eel state
show such degeneracy at four wavenumbers surrounding
${\bf Q} = (\pi/a)({\bf{\hat x}}+{\bf{\hat y}})$ along the principal axes.
Spin resonances in superconducting FeSe intercalates have been observed
recently at these wavenumbers by inelastic neutron scattering~\cite{davies_16,pan_16,ma_16}.
This suggests that hidden N\'eel order is present in heavily electron-doped FeSe.

\section{One-Electron/One-Hole Bands}
\begin{figure}
\centering
\includegraphics[scale=0.50, angle=-90]{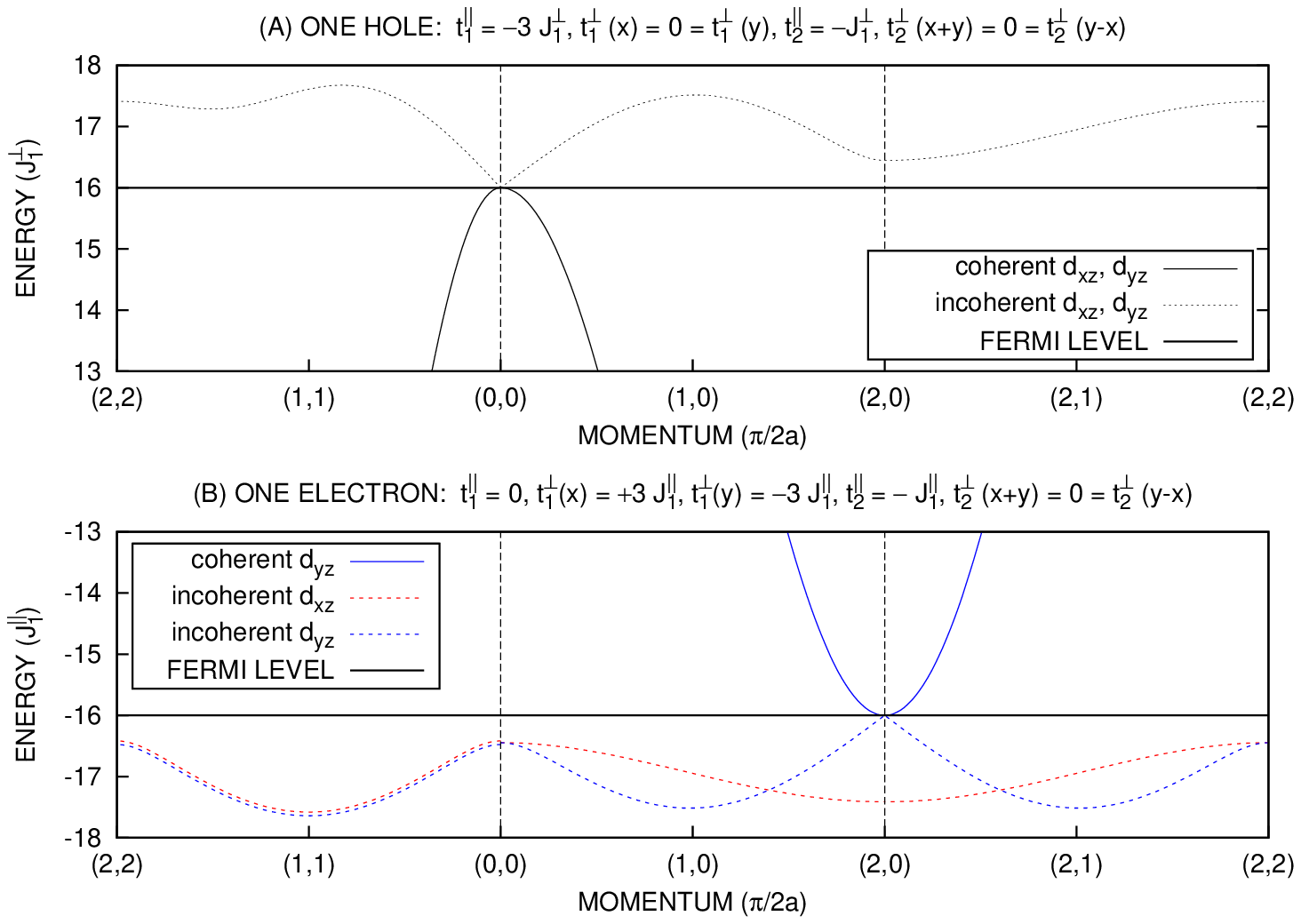}
\caption{One-electron spectra of hidden half metal states
within the mean field approximation
at site-orbital concentration approaching half filling, $x \rightarrow 0$
(see refs.~\cite{jpr_16,jpr_17}).
Heisenberg exchange coupling constants are set in
the caption to Figure~\ref{spin_waves}, while
$J_0 = J_{0c} + 0.1\, J_1^{(\perp)\parallel}$. }
\label{spectra}
\end{figure}

We   now compare 
spectra for one mobile hole and for one mobile electron
with respect to half filling,
with $t$-$J$ model parameters that are related to each other 
by the previous particle--hole transformation in Equation~(\ref{tJbar1}).  
In the hole-doped case, 
the Heisenberg exchange coupling constants 
coincide with the previous set for the hidden ferromagnet (Figure \ref{spin_waves}a),
while the hopping matrix elements are set to
$t_1^{\parallel} = -3\, J_1^{\perp}$,
$t_1^{\perp}({\bf{\hat x}}) = -2\,J_1^{\perp}$,
$t_1^{\perp}({\bf{\hat y}}) = +2\,J_1^{\perp}$,
$t_2^{\parallel} = -J_1^{\perp}$ and $t_2^{d\pm,d\mp}=0$.  
The latter implies that
the $d_{xz}$ (even) versus the $d_{yz}$ (odd) orbital
is a good quantum number.
In the electron-doped case,
nearest neighbor $t$-$J$ model parameters 
are set by Equation~(\ref{tJbar1}),
while on-site and  next-nearest neighbor model parameters are unchanged.
Figure~\ref{no_hunds_rule} shows exact spectra
for one mobile hole and for one  mobile electron
roaming over a periodic $4\times 4$ lattice of iron atoms,
in the absence of Hund's Rule, $J_0 = 0$.
The Schwinger-boson-slave-fermion representation of the correlated electron (hole)
in the limit $U_0\rightarrow\infty$
was exploited in such case~\cite{kane_89,auerbach_larson_91}.
Details of the numerical calculation are given 
in the Supplementary Materials and in Ref.~\cite{jpr_mana_pds_14}.
Notice that all of the states obey the particle--hole transformation in Equation~(\ref{p_to_h}).
Figure~\ref{spectra} shows the one-electron spectra predicted by Schwinger-boson-slave-fermion
mean field theory, but at $t_1^{\perp} = 0$ for hole doping,
and at $t_1^{\parallel} = 0$ for electron doping.
Notice that the states again obey the particle--hole relationship in Equation~(\ref{p_to_h})\
depicted by Figure~\ref{e_strctr}.

The dispersion of the lowest-energy spin-1/2 mobile-hole states shown by
Figure~\ref{no_hunds_rule}a can be understood at ideal hopping, 
achieved by suppressing nearest-neighbor inter-orbital hopping as well: 
$t_1^{\perp} \rightarrow 0$.  
A half metal characterized by hidden magnetic order
depicted by the inset to Figure~\ref{no_hunds_rule}a
is predicted in the absence of Hund's Rule
at large electron spin $s_0$~\cite{jpr_mana_pds_11,jpr_mana_pds_14}.  
Electrons are spin polarized per $d\pm$ orbital,
where they follow a hole-type energy dispersion relation
$\varepsilon_e^{(0)} ({\bf k}) = -2t_1^{\parallel}\sum_{n=x,y} \cos\,k_n a
-2t_2^{\parallel}\sum_{n=+,-} \cos\,k_n a $,
with $k_{\pm} = k_x\pm k_y$
(cf.    the {\it true} half metal in ref.~\cite{MLT}).
Two degenerate hole Fermi surface pockets at zero 2D moment are predicted 
for small concentrations of mobile holes per orbital, $x$,
each with a Fermi wavenumber $k_F a = (4\pi x)^{1/2}$
(see Figure~\ref{e_strctr}a).
The top of the hole-type band lies
$\epsilon_F = |t_1^{\parallel} + 2 t_2^{\parallel}| (k_F a)^2$
above the Fermi level.
These coherent hole bands are recovered by a calculation of
the one-electron propagator 
within a Schwinger-boson-slave-fermion mean-field approximation~\cite{kane_89,auerbach_larson_91}
of the two-orbital $t$-$J$ model in Equation~(\ref{tJ})
for the above hidden half metal~\cite{jpr_mana_pds_11,jpr_mana_pds_14}.
In the limit near half filling,
at $|t|\gg J$, 
the one-electron propagator also reveals composite electron--spin--wave states
at an energy 
$\epsilon_F + \hbar\omega_{\rm sw}({\bf k})$ above the Fermi level, where
$\omega_{\rm sw}({\bf k})$ is the spin-wave dispersion
at large electron spin $s_0$ shown 
by Figure~\ref{spin_waves}a,b~\cite{jpr_10} 
(see also Figure~\ref{spectra}a).
They are  incoherent excitations that show intrinsic broadening~\cite{jpr_16}.
The predicted dispersion relation
is traced by the dashed line in Figure~\ref{no_hunds_rule}a in the absence of Hund's Rule.
It compares well with the exact dispersion of the lowest-energy spin-1/2
excitations at non-ideal hopping matrix elements,
in the absence of Hund's Rule, 
and it
notably shows electron-type dispersion in the vicinity of cSDW wavenumbers
$(\pi/a){\bf{\hat x}}$ and $(\pi/a){\bf{\hat y}}$.
The latter 
are pulled down to lower energy as Hund coupling is turned on 
(cf. Figure~S2a in Supplementary Materials).
We~therefore interpret the dispersion of
those spin-1/2 groundstates,
which, respectively, have odd and even parity under orbital swap $P_{d,{\bar d}}$,
as emergent $d_{yz}$ and $d_{xz}$ electron bands.

Application of the particle--hole transformation in Equation~(\ref{spc}) 
yields a new hidden half metal state
depicted by the inset to Figure~\ref{no_hunds_rule}b, 
where the missing spin-1/2 moment 
in the third row represents a spin singlet  
(cf.   ~the {\it true} half metal in Ref.~\cite{MLT}).
By  Equation~(\ref{tJbar1}),
it is governed by the two-orbital $t$-$J$ model in Equation~(\ref{tJ})
at electron doping above half-filling,
with Heisenberg coupling constants that coincide with the previous set
for the hidden N\'eel state (Figure~\ref{spin_waves}c),
and with hopping parameters
$t_1^{\parallel} = +2\, J_1^{\parallel}$,
$t_1^{\perp}({\bf{\hat x}}) = +3\,J_1^{\parallel}$,
$t_1^{\perp}({\bf{\hat y}}) = -3\,J_1^{\parallel}$,
$t_2^{\parallel} = -J_1^{\parallel}$, and 
$t_2^{d\pm,d\mp} = 0$.
As $t_1^{\parallel}\rightarrow 0$,
Schwinger-boson-slave-fermion mean field theory
applied to the new model
predicts circular electron Fermi surface pockets at cSDW wavenumbers
similar to Figure~\ref{e_strctr}b.  
It also predicts emergent {\it hole} excitations
that disperse according to
the dashed lines in Figure~\ref{no_hunds_rule}b~\cite{jpr_17}
(see also Figure~\ref{spectra}b). 
Again, the exact spectrum compares well to mean field theory.
The first and second excited spin-1/2 states in Figure~\ref{no_hunds_rule}b
that lie at momentum zero and $(\pi/a)({\bf{\hat x}}+{\bf{\hat y}})$
thereby correspond to a hole band plus its replica at lower energy,
both buried below the Fermi level at zero 2D momentum 
in the two-iron folded Brillouin zone.
Turning on Hund coupling pulls the first excited state
at zero 2D momentum down in energy
(cf. Figure~S2b in Supplementary Materials). 
The~previous prediction is consistent
with reported evidence for such 
a replica band at the $\Gamma$ point
from ARPES on FeSe/STO~\cite{lee_14}.

\section{Cooper Pairs with Emergent Sign Changes}
Consider now
two electrons above half filling 
that roam over a $4\times 4$ periodic lattice of iron atoms 
governed by the two-orbital $t$-$J$ model in Equation~(\ref{tJ}) \cite{jpr_17}.
Heisenberg exchange parameters
are set to those listed in the caption to Figure~\ref{spin_waves}c,
but new hopping matrix elements are chosen
that leave the electron masses $m_x$ and $m_y$ per orbital
unchanged at cSDW momenta:
$t_1^{\parallel} = 2\, J_1^{\parallel}$,
$t_1^{\perp}({\bf{\hat x}}) = +5\, J_1^{\parallel}$, 
$t_1^{\perp}({\bf{\hat y}}) = -5\, J_1^{\parallel}$,
and $t_2^{\alpha,\beta} = 0$.
Such model parameters result in electron-type Fermi surface pockets
centered at cSDW momenta
in the hidden half metal state within the mean field approximation.
Details of the exact calculation are given in
the Supplementary Materials and in Ref.~\cite{jpr_16}.
The Hund coupling, $-J_0$, is tuned to a
putative
quantum critical point (QCP)
defined by degeneracy of the spin resonance at cSDW momenta 
with the hidden spin resonance 
at momentum $(\pi/a)({\bf{\hat x}}+{\bf{\hat y}})$.
This definition is suggested by
the semi-classical analysis of the corresponding Heisenberg model at half filling (Figure~\ref{spin_waves}c),
which finds a QCP
when the spin gap at cSDW momenta collapses~\cite{jpr_10}: 
$\Delta_{cSDW}\rightarrow 0$.
A bound electron-pair groundstate exists below a continuum of states
at zero net 2D momentum.
It shows $S$-wave symmetry
according to the reflection parities listed in Table~\ref{table1}.
An excited pair state with $D_{x^2-y^2}$
symmetry exists below the continuum as well.

\begin{table}
\caption{Reflection parities, 
orbital-swap parity,
and spin of low-energy pair states with zero net momentum
in order of increasing energy.  
The operator $R_{x^{\prime}z}$, for example,
denotes a reflection about the $x^{\prime}$-$z$ plane,
where $x^{\prime}$ is a diagonal axis.
The hidden spin--wave in the case of electron doping is exceptional,
where it carries net momentum $(\pi/a)({\bf{\hat x}}+{\bf{\hat y}})$.}
\label{table1}
\centering
\begin{tabular}{cccccc}
\hline
No. & Pair/Particle-Hole State & $R_{xz}$ , $R_{yz}$ & $R_{x^{\prime}z}$ , $R_{y^{\prime}z}$ & $P_{d,{\bar d}}$ & Spin \\
\hline
$0$  & $S$ & $+$ & $+$ & $+$ & $0$  \\
$1$ & $D_{x^2-y^2}$ & $+$ & $-$ & $+$ & $0$ \\
$2$ & hidden  spin--wave & $-$ & $-$ & $-$ & $1$ \\
\hline
\end{tabular}
\end{table}

\begin{figure}
\centering
\includegraphics[scale=0.88]{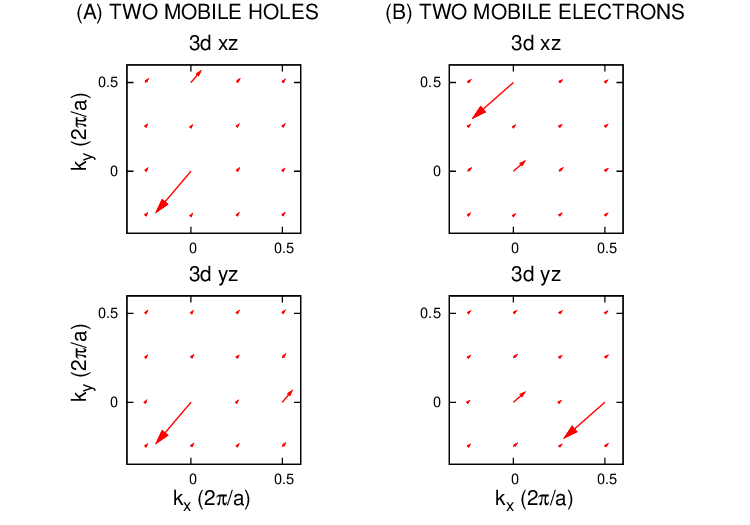}
\caption{Complex order parameter for superconductivity (Equation~(\ref{iF})), with
Heisenberg exchange coupling constants set in the caption to Figure~\ref{spin_waves},
and with hopping matrix elements (\textbf{a})
$t_1^{\parallel} = -5\, J_1^{\perp}$,
$t_1^{\perp}({\bf{\hat x}}) = -2\, J_1^{\perp}$,
$t_1^{\perp}({\bf{\hat y}}) = +2\, J_1^{\perp}$,
and $t_2^{\alpha,\beta} = 0$ for two mobile holes ($30$ electrons).
Nearest neighbor hopping matrix elements transform to (\textbf{b}) 
$t_1^{\parallel} = 2\, J_1^{\parallel}$,
$t_1^{\perp}({\bf{\hat x}}) = +5\, J_1^{\parallel}$ and
$t_1^{\perp}({\bf{\hat y}}) = -5\, J_1^{\parallel}$
and $t_2^{\alpha,\beta} = 0$
for two mobile electrons ($34$ electrons).
In addition, inter-orbital on-site repulsion is set to
$U_0^{\prime} = {1\over 4}J_0 + 1000\, J_1^{(\perp)\parallel}$,
while the Hund coupling constant is set to
$-J_0 = 2.25\, J_1^{(\perp)\parallel}$.
Heisenberg-exchange interactions in the Hamiltonian in Equation~(\ref{tJ}) 
are replaced with 1/2 the corresponding 
spin-exchange operators to reduce finite-size~effects.}
\label{30_qcp}
\end{figure}

The order parameter for superconductivity is the defined as
\begin{equation}
iF(k_0,{\bf k}) = \langle \Psi_{\rm Mott}|
{\tilde c}_{\uparrow}(k_0,{\bf k})
{\tilde c}_{\downarrow}(k_0,-{\bf k})|\Psi_{\rm Cooper}\rangle
\label{iF}
\end{equation}
times $\sqrt 2$, where 
$|\Psi_{\rm Cooper}\rangle$
is the groundstate of the electron pair,
and where $\langle \Psi_{\rm Mott}|$ is the 
groundstate of the  Mott insulator at half filling.
Above, the tilde notation signals the limit $U_0\rightarrow\infty$.
Figure~\ref{30_qcp}b depicts  Equation~(\ref{iF})
using exact groundstates
$\langle \Psi_{\rm Mott}|$ and $|\Psi_{\rm Cooper}\rangle$
on a $4\times 4$ periodic lattice of iron atoms
at the putative QCP.
In particular,
the Hund coupling is tuned so that 
the groundstate spin-$1$ states at cSDW momenta,
which have even parity under orbital swap, $P_{d,{\bar d}}$, 
become degenerate with the groundstate spin-$1$ state
at wavenumber $(\pi/a)({\bf{\hat x}}+{\bf{\hat y}})$,
which has odd parity under orbital swap.
The~coupling constants, respectively, are
$-J_0 = 1.35\, J_1^{\parallel}$ and $-J_0 = 2.25\, J_1^{\parallel}$
at half filling and for two mobile electrons.
Notice that the order parameter displayed by Figure~\ref{30_qcp}b
is isotropic, but that it alternates in sign
between the emergent hole bands at zero 2D momentum and the electron bands at
cSDW momenta~\cite{jpr_17}.

Figure~\ref{30_qcp}a shows the particle--hole conjugate 
of the order parameter in Equation~(\ref{iF}) for superconductivity
in the two-orbital $t$-$J$ model with two-mobile holes
that roam over a $4\times 4$ periodic lattice,
under the transformation in Equation~(\ref{tJbar1})
in parameter space~\cite{jpr_16}.
Notice that it is related
to Figure~\ref{30_qcp}b by the particle--hole transformation in Equation~(\ref{p_to_h}).
In conclusion, both the electron pair and the conjugate hole pair 
display an $S^{+-}$ order parameter for superconductivity, 
with remnant pairing on the emergent band of opposite sign.
This result is similar to
a recent proposal for 
$S^{+-}$ pairing in heavily hole-doped iron superconductors
that is based on
a phenomenological attractive pairing interaction~\cite{bang_14}.

\section{Discussion and Conclusions}
Heavily electron-doped surface layers of FeSe show record superconducting critical
temperatures as high as $T_c \cong 100$ K~\cite{ge_15}.
ARPES reveals
two electron Fermi-surface pockets at the corner of the two-iron Brillouin zone
that cross, and that do not show level repulsion~\cite{peng_14}.
The electronic structure at the surface layer of heavily electron-doped  FeSe
can be described by the two-orbital $t$-$J$ model in Equation~(\ref{tJ}) 
at sub-critical Hund coupling,
with hopping matrix elements and Heisenberg exchange coupling constants 
that favor the half metal state shown in the inset to
Figure~\ref{no_hunds_rule}b.
In particular, exact results and
Schwinger-boson-slave-fermion mean field theory 
predict electron Fermi surface pockets
centered at the two distinct cSDW momenta following Figure~\ref{e_strctr}b 
(see Figure~S2b in Supplementary Materials and Figure~\ref{spectra}b).
The~Cooper pairs in heavily electron-doped FeSe surface layers
are isotropic at the electron pockets~\cite{peng_14,lee_14,zhao_16},
but we propose that they change sign at the buried hole bands
according to Figure~\ref{30_qcp}b~\cite{jpr_17}.

Application of the particle--hole
transformation in Equation~(\ref{spc}) to the two-orbital $t$-$J$ model
for a surface layer of FeSe implies
a high-$T_c$ surface layer at heavy hole doping 
that shows hole-type Fermi surface pockets 
at the center of the Brillouin zone
(Figure~\ref{e_strctr}a).
It suggests searching for high-$T_c$ superconductivity
in surface layers of hole-doped iron-based compounds.  

Finally,
from a purely technical perspective,
the particle--hole transformation in Equation~(\ref{spc})
of the two-orbital Hubbard model in Equation~(\ref{tJ})
for iron-based superconductors 
is a valuable tool that helps map out the parameter space.
Figures~\ref{no_hunds_rule} and \ref{spectra} explicitly confirm 
the validity of the particle--hole transformation
in the case of exact diagonalization on finite clusters and in the case of the
mean-field approximation of the Schwinger-boson-slave-fermion formulation. 
The particle--hole transformation in Equation~(\ref{spc})
will play a useful role in future analyses of the two-orbital Hubbard model in Equation~(\ref{tJ})
for iron-based superconductors by other techniques,
such as by quantum Monte Carlo~\cite{hirsch_85},
and by experimental simulations using trapped atoms~\cite{casanova_12}
and superconducting qubits~\cite{barends_15}.

\vspace{6pt}

The author thanks Miguel Araujo for correspondence.  He also thanks
Brent Andersen, Richard Roberts and Timothy Sell for
technical help with the use of
the virtual shared-memory cluster (Lancer) at
the AFRL DoD Supercomputing Resource Center.
This work was supported in part by the US Air Force
Office of Scientific Research under grant Nos. FA9550-13-1-0118 and FA9550-17-1-0312,
and by the National Science Foundation under PREM grant No. DMR-1523588.


\appendix
\section{Two-Orbital Particle--Hole Transformation}\label{app}
Below, we derive the particle--hole transformation for iron superconductors
formulated in momentum space (Figure~\ref{e_strctr}) and Equation~(\ref{p_to_h})),
starting from the transformation in real space in Equation~(\ref{spc}):
\begin{equation}
c_{i,\alpha,s}\rightarrow (-1)^{y_i/a} c_{i,p_i(\alpha),s}^{\dagger}
\quad {\rm and} \quad
c_{i,\alpha,s}^{\dagger}\rightarrow (-1)^{y_i/a} c_{i,p_i(\alpha),s},
\label{s_spc}
\end{equation}
%
where $p_i(d\pm) = d\pm$ for iron sites $i$ on the $A$ sublattice of the checkerboard,
and where $p_i(d\pm) = d\mp$ for iron sites $i$ on the $B$ sublattice of the checkerboard.
The creation operator for a spin $s$ electron that carries $3$-momentum $(k_0,{\bf k})$
is
\begin{equation}
c_s^{\dagger}(k_0,{\bf k}) = {\cal N}^{-1/2}\sum_{\alpha = 0}^1\sum_i
e^{i(k_0 \alpha + {\bf k}\cdot{\bf r}_i)} c_{i,\alpha,s}^{\dagger},
\label{s_mmntm_spc}
\end{equation}
where
${\cal N} = 2 N_{\rm Fe}$ denotes the number of sites-orbitals on the square lattice of iron atoms,
and where the indices $0$ and $1$ denote
the $d-$ and $d+$ orbitals $\alpha$, respectively.
The quantum numbers $k_0 = 0$ and $\pi$ therefore represent
the $d_{xz}$ and the $(-i)d_{yz}$ orbitals.

Following Equation~(\ref{s_mmntm_spc}),
taking the Fourier transform of the first particle--hole transformation
in Equation~(\ref{s_spc})
therefore yields that the destruction operator
$c_s(k_0,{\bf k})$ transforms~to
\begin{equation}
{\cal N}^{-1/2}\sum_{\alpha = 0}^1\sum_i
e^{i(k_0 \alpha + {\bf k}\cdot{\bf r}_i)}  (-1)^{y_i/a} c_{i,p_i(\alpha),s}^{\dagger}
= {\cal N}^{-1/2}\sum_{\alpha = 0}^1\sum_i
e^{i[k_0 \alpha + ({\bf k}+{\pi\over a}{\bf{\hat y}})\cdot{\bf r}_i ]}
 c_{i,p_i(\alpha),s}^{\dagger},
\label{s_mmntm_spc_trnfm}
\end{equation}
which is explicitly
\begin{equation}
{\cal N}^{-1/2}\sum_{\alpha = 0}^1\Biggl\{\sum_{i\in A}
e^{i(k_0 \alpha + ({\bf k}+{\pi\over a}{\bf{\hat y}})\cdot{\bf r}_i)}  c_{i,\alpha,s}^{\dagger}
+\sum_{i\in B}
e^{i[k_0 (\alpha+1) + ({\bf k}+{\pi\over a}{\bf{\hat y}})\cdot{\bf r}_i ]}
 c_{i,\alpha,s}^{\dagger}\Biggr\}.
\label{s_mmntm_spc_trnfm_prm}
\end{equation}
Comparison with Equation~(\ref{s_mmntm_spc}) yields that the above coincides with
$c_s^{\dagger}(k_0,{\bf k}+{\bf Q}_{k_0})$
 at $k_0 = 0$,
where ${\bf Q}_{0} = (\pi/a) {\bf{\hat y}}$.
At $k_0 = \pi$, on the other hand, the extra factor of
$e^{i k_0} = -1$ in the second term of Equation~(\ref{s_mmntm_spc_trnfm_prm})
can be replaced by an {\it overall} factor of
$e^{i{\pi\over a}({\bf{\hat x}}+{\bf{\hat y}})\cdot{\bf r_i}}$.
Comparison with Equation~(\ref{s_mmntm_spc}) in turn yields that Equation          (\ref{s_mmntm_spc_trnfm_prm})
coincides with $c_s^{\dagger}(k_0,{\bf k}+{\bf Q}_{k_0})$ at $k_0 = \pi$,
where ${\bf Q}_{\pi} = (\pi/a) {\bf{\hat x}}$.  We thereby obtain the particle--hole
transformation in momentum space:
$c_s(k_0,{\bf k}) \rightarrow c_s^{\dagger}(k_0,{\bf k}+{\bf Q}_{k_0})$.
Using the identity ${\bf Q}_{k_0} + {\bf Q}_{k_0} = 0$
that is true for crystal momentum
yields the conjugate particle--hole
transformation in momentum space:
$c_s^{\dagger}(k_0,{\bf k}) \rightarrow c_s(k_0,{\bf k}+{\bf Q}_{k_0})$.


\end{document}